\documentstyle[aps,multicol]{revtex}
\begin{document}
\draft
\title
{
Gaussian field theories,
random Cantor sets and
multifractality
}

\author{Claudio de C. Chamon, Christopher Mudry and Xiao-Gang Wen}
\address{Department of Physics, Massachusetts Institute of Technology,
77 Massachusetts Avenue, Cambridge, Massachusetts 02139}
\date{\today}
\maketitle

\begin{abstract}
The computation of multifractal scaling properties associated
with a critical field theory involves non-local operators and remains
an open problem using conventional techniques of field theory.
We propose a new description of Gaussian field theories
in terms of random Cantor sets and show how universal multifractal scaling
exponents can be calculated.
We use this approach to characterize the
multifractal critical wave function of Dirac fermions interacting with a
random vector potential in two spatial dimensions.

\end{abstract}



\begin{multicols}{2}
\narrowtext

Studies of critical theories
are usually concentrated on scaling properties of {\it local} operators and
their correlations.
However, there are objects in critical systems (at least in critical
theories of disordered systems)
that have very complex scaling properties, namely {\it multifractal}
\cite{Mandelbrot 1974,Hentschel 1983,Frisch 1985,Halsey 1986}.
They are characterized not by a single scaling exponent, but by
infinitely many scaling dimensions.
An open question of field theory is how to describe such
complex scaling behavior, in particular how to calculate the
multifractal scaling functions $\tau(q)$ or  $f(\alpha)$
\cite{Kadanoff 1990,Duplantier 1991,Deutsch 1994}.

Multifractality is a property of a distribution in a critical theory
or some other deterministic systems, such as dynamical
systems \cite{Hentschel 1983,Halsey 1986}.
The key feature of the phenomenon of multifractality is
the property of selfsimilarity of the distribution at different length scales.
A more refined description of the property
of selfsimilarity relies on the concept of box distribution.
Let $\Omega$ be a large box of volume $L^d$,
and let $\mu(x)$ be a {\it normalized} distribution
(assumed to be strictly positive) defined on $\Omega$.
Imagine dividing $\Omega$ into
$M$ boxes $\Omega^{\ }_i$, each of volume $a^d$.
The box distribution is then defined by
$
\{p^{\ }_i=
\int_{\Omega^{\ }_i}\!\! d^dx\ \mu(x),
\ i=1,\dots,M
\}
$,
and the distribution $\mu(x)$ is selfsimilar
if the scaling law
\begin{equation}
\tau(q)\equiv D(q)(q-1)=
\lim_{{a\over L}\rightarrow0}
{1\over\ln({a\over L})}
\ln\left(\sum_{i=1}^M p^q_i\right)
\label{exacttau(q)}
\end{equation}
holds for real $q$
\cite{Hentschel 1983}.  The
function $\tau(q)$ can be shown to be concave and strictly increasing
\cite{Hentschel 1983}. If $D(q)$ is constant,
the distribution $\mu(x)$ is said to be simple fractal of $D$ (Hausdorff)
dimensional support. If $D(q)$ is not constant [$\tau(q)$ nonlinear],
the distribution $\mu(x)$ is {\it multifractal}.

The scaling function $\tau(q)$ is known for many examples of
deterministic multifractals \cite{Hentschel 1983,Halsey 1986}.
However, much less is known for distributions in critical statistical
systems. Statistical multifractals can be found in the problem of
localization in two spatial dimensions where the distribution $\mu(x,V)$
is constructed from the squared amplitude of the wave function solving
Schr\" odinger equation with an impurity potential $V(x)$
\cite{Castellani 1986}. In this context,
a very important example of multifractality
is that of the critical wave function at the plateau
transition in the integer quantum Hall effect (IQHE).
Numerical calculations of $\tau^{\ }_{\rm iqhe}(q)$
for the IQHE have been performed for different realizations
of the impurity potential. Within numerical errors,
$\tau^{\ }_{\rm iqhe}(q)$ is independent of the disorder \cite{Pook 1991}.
Moreover, the exponent $\tau^{\ }_{\rm iqhe}(2)$, which
governs anomalous diffusion \cite{Chalker 1988},
is consistent with the value $\nu\approx7/3$
for the localization length exponent \cite{Huckestein 1994}.
Very little is known on the analytical
structure of the critical theory
in the IQHE. Much more is known, however, on that for
Dirac fermions in two spatial dimensions
interacting with the random vector potential
$
A^{\ }_{\nu}(x)=
$
$
\epsilon^{\ }_{\nu\rho}\partial^{\ }_{\rho}\Theta(x)+
                      \partial^{\ }_{\nu}\chi(x),
$
distributed according to
$\exp[-{1\over2g^{\ }_A}\int\! d^2x A^2_{\nu}]$.
In this problem of localization, one can construct a distribution
\begin{eqnarray}
\mu(x,\Theta)\!\propto\!
{e^{2\Theta(x)}\over\int\! d^2x\ e^{2\Theta(x)}},
\
P[\Theta]\!\propto\!
e^{-{1\over2g^{\ }_A}\!\int\! d^2x (\nabla\Theta)^2},
\label{Diracprobdistr}
\end{eqnarray}
from the zero-energy eigenfunctions
$\psi(x)\propto\exp[\Theta(x)+i\chi(x)]$
\cite{Ludwig 1994}.
Note that in Eq. (\ref{Diracprobdistr}) we are dealing with a {\it
statistical ensemble} of distributions with weights $P[\Theta]$.
Although the disordered critical point is exactly solvable \cite{Ludwig
1994}, we still lack a valid calculation of $\tau^{\ }_{\rm Dirac}(q)$
for the distributions in Eq. (\ref{Diracprobdistr}).
This is so due to the failure of the crucial assumption commonly used
in field theoretical calculations of multifractal scaling exponents
\cite{Mudry 1995},
whereby it is assumed that averaging $e^{2q\Theta(x)}$
over space is equivalent to averaging over the disorder
\cite{Duplantier 1991,Ludwig 1994}.
Under this assumption, one may replace the inverse of the normalization factor
$\int d^2 x e^{2\Theta(x)}$ by a number, and use
conventional methods from field theory. But, the resulting
$\tau^*_{\rm Dirac}(q)$
does not satisfy the analytical properties of Eq. (\ref{exacttau(q)}).
When  $e^{2\Theta(x)}$ is not selfaveraging,
conventional methods
can not treat random variables such as
$[\int d^2 x e^{2\Theta(x)}]^{-1}$
anymore.

One goal of this letter is to calculate $\tau^{\ }_{\rm Dirac}(q)$
exactly for the first time. We hope that by doing so, we can gain some
familiarity with the unusual
properties of critical phenomenon for disordered system,
in particular those associated with the plateau transition in the IQHE.
Since conventional methods
relying on the underlying critical field theory are unsatisfactory
\cite{Mudry 1995}, we take a very different approach. We first construct
explicitly a model for a statistical ensemble of distributions
$\mu(x,\Phi)$ ($\Phi$ labels the members of the ensemble) and then
show that this model is closely related to the critical theory of
Dirac fermions with random vector potentials as described, say, in
\cite{Mudry 1995}.
Our model is a simple generalization of the
deterministic two scale Cantor set multifractal
\cite{Hentschel 1983,Halsey 1986}.
Hence, we call it a random Cantor set (RCS) construction.  It should
be said that the model itself has been used, for example, in the study
of turbulences \cite{Mandelbrot 1974} and of directed polymers in a
random medium \cite{Derrida 1988}.  The novelty of our work is to
relate the RCS to the Gaussian field theory
${\cal L}={1\over2}(\nabla\Phi)^2$,
and then use this relationship to calculate
$\tau^{\ }_{\rm Dirac}(q)$.
In this way, we hope to illustrate how a new type of equivalence
between critical field theories can be taken advantage of.

Our first result is to show that a particular RCS construction
describes a Gaussian field theory on an ultrametric space. The
correlation functions share the same form as that in the theory
${\cal L}={1\over2}(\nabla\Phi)^2$ with an Euclidean metric.
Although the two metrics differ, scaling exponents of correlation functions
are the same in the two theories. Thus, we can use the one
realization in which the calculation is simpler. In the case of the
multifractal scaling exponents $\tau(q)$, the RCS is the
realization of choice.

Using the RCS realization, we can then show that almost all members
$\mu(x,\Phi)$ of the statistical ensemble yield, with the help of Eq.
(\ref{exacttau(q)}), a scaling exponent $\tau(q,\Phi)$ which is
selfaveraging, {\it i.e.}, is almost surely independent of $\Phi$ in
the limit ${a\over L}\rightarrow0$. To derive this result, we borrow
an exact solution for the free energy of directed polymers on a
random tree showing that the free energy is selfaveraging
\cite{Derrida 1988,Buffet 1993}.

We provide support for the claim that the scaling exponents $\tau(q)$
should be the same in both theories by comparing exact results from
the RCS construction with numerical simulations for the
${\cal L}={1\over2}(\nabla\Phi)^2$ theory.
We also present analytical arguments which indicate a phase
transition in $\tau(q)$ for the ${\cal L}= {1\over2}(\nabla\Phi)^2$
theory at exactly the same $q^{\ }_c$ for which a transition occurs in
the RCS model.

\begin{figure}
\begin{picture}(60,60)(110,360)
\thicklines
\put(230,420){\line( 1,-1){ 30}}\put(190,405){$\ \phi_{11}$}
\put(200,390){\line( 1,-1){ 20}}\put(170,380){$\phi_{21}$}
\put(200,390){\line(-1,-1){ 20}}\put(210,380){$\phi_{22}$}
\put(230,420){\line(-1,-1){ 30}}\put(250,405){$\ \phi_{12}$}
\put(260,390){\line( 1,-1){ 20}}\put(230,380){$\phi_{23}$}
\put(260,390){\line(-1,-1){ 20}}\put(270,380){$\ \phi_{24}$}
\end{picture}
\caption
{
First two levels of RCS model.
}
\label{binarytree}
\end{figure}
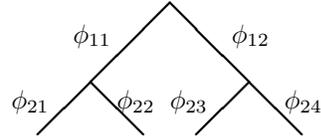

We begin with the construction of the RCS model.  Consider a box
$\Omega$ of volume $L^2$ in two dimensional Euclidean space.
Imagine dividing $\Omega$ into two boxes of equal volume $L^2/2$
and assign to each half a binary address $s^{\ }_1=\pm$, or, equivalently,
the address $(1,1)$ and $(1,2)$.
The procedure is then iterated $n$ times.
At level $n$ the space is divided into $2^n$
small boxes addressed by either $n$ bits $(s^{\ }_1,\dots,s^{\ }_n)$ taking
the values $s^{\ }_i=\pm$, $i=1,\dots,n$ or by $(n,j)$ where $j=1,\dots,2^n$.
Any arbitrary point
$x\in\Omega$ is then uniquely addressed according to the box in which
it lies, say, by the binary address $(s^x_1,\dots,s^x_n)$.
Having constructed a selfsimilar structure, namely a binary tree, we
define on it random variables.  At each level or generation
$i=1,\dots,n$ of the binary tree one draws $2^i$ independent random
variables $\phi^{\ }_{s^{\ }_1\dots s^{\ }_i}\equiv\phi^{\ }_{ij}$,
from a Gaussian probability distribution $P[\phi]=({1\over2\pi
g})^{1/2}\exp(-{\phi^2\over2g})$
(see Figure \ref{binarytree}).
Finally, for any
given binary tree made of $n$ generations, we define the random
variable
\begin{eqnarray}
\Phi^{\ }_n(x)=
\sum_{i=1}^n \phi^{\ }_{s^x_1\dots s^x_i}=
\sum_{i=1}^n\left[\sum_{j=1}^{2^i}e^{\ }_{ij}(x^{\ }_k)\phi^{\ }_{ij}\right].
\label{Phin(x)}
\end{eqnarray}
Here, for any generation $i$ there exists one and only one integer
$0\leq j_i(x^{\ }_k)\leq 2^i$ such that
$e^{\ }_{ij}(x^{\ }_k)=\delta^{\ }_{jj_i(x^{\ }_k)}$.
By construction, $\Phi^{\ }_n(x)$ is a random Gaussian variable.
We see that random variables $\phi^{\ }_{s^x_1\dots s^x_i}$ at different
levels $i$ describe fluctuations at different length scales. Thus,
$\Phi_n(x)$, containing fluctuations at all length scales, has a
selfsimilar structure.

Next, we calculate the generating function of $N$-point correlation
functions for the RCS model [Eq. (\ref{Phin(x)})].
We show that it is quadratic in the sources, {\it i.e.}, describes a
Gaussian field theory. We then discuss the relationship between the
Gaussian field theory for the RCS model and the
Euclidean field theory ${\cal L}= {1\over2}(\nabla\Phi)^2$.

We consider an arbitrary collection of $N$ points $x^{\ }_1,\dots,x^{\
}_N$ in a large box $\Omega$ of volume $L^2$ embedded in two
dimensional Euclidean space.  We want to calculate the correlation
function
\begin{equation}
{\cal Z}^{\ }_n(1,\dots,N)\equiv
\overline{\exp\left[i\sum_{k=1}^N q^{\ }_k\Phi^{\ }_n(x^{\ }_k)\right]},
\end{equation}
where $q^{\ }_1,\dots,q^{\ }_N$ are real and the overline denotes averaging
over all random variables $\phi^{\ }_{ij}$.
The Gaussian random variables $\Phi^{\ }_n(x^{\ }_k)$, $k=1,\dots,N$,
are not necessarily independent.
Their statistical correlations can be measured by the scalar product
$I^{\ }_{kl}\equiv$
$
\sum_{i=1}^n
[\sum_{j=1}^{2^i}e^{\ }_{ij}(x^{\ }_k)e^{\ }_{ij}(x^{\ }_l)].
$
$I^{\ }_{kl}$
measures how ``close'' $x^{\ }_k$ and $x^{\ }_l$ are in the binary
partition of $\Omega$ by returning the number $I^{\ }_{kl}$ of common
boxes of volume $L^2/2,\dots,L^2/2^{-I^{\ }_{kl}}$,
respectively, that contain both $x^{\ }_k$ and $x^{\ }_l$.
On the other hand,
one can perform the Gaussian
integrations over all independent Gaussian random variables
$\phi^{\ }_{ij}$, and one finds
\begin{equation}
{\cal Z}^{\ }_n(1,\dots,N)= e^{-{ng\over2}\sum_{k=1}^N q^2_k -g\sum^{\
}_{1\leq k<l\leq N} q^{\ }_k I^{\ }_{kl} q^{\ }_l}.
\label{generatingf}
\end{equation}
Eq. (\ref{generatingf})
shows that the RCS model defined by Eq. (\ref{Phin(x)})
describes a Gaussian field theory, since Eq. (\ref{generatingf}) implies that
$N$-point functions are constructed from 2-point functions in the same way
as in the ${\cal L}= {1\over2}(\nabla\Phi)^2$ theory.

Now, let $d(x_k,x_l)=L\ 2^{-I^{\ }_{kl}/2}$. One can easily show that
$d(x_k,x_m)\leq{\rm max}[d(x_k,x_l),d(x_l,x_m)]$,
$\forall x_k,x_l,x_m\in\Omega$,
{\it i.e.}, $d(x,y)$ is an ultrametric (and thus a metric),
and that the 2-point function in the binary tree is given by
\begin{equation}
G_{\rm RCS}(x,y)\equiv
\overline{\Phi^{\ }_n(x) \Phi^{\ }_n(y)}=-{2g\over\ln2}
\ln {d(x,y)\over L}.
\label{greenRCS}
\end{equation}
Thus, the RCS 2-point function resembles very much the 2-point function
$G(x,y)=-{g^{\ }_A\over2\pi}\ln{|x-y|\over L}$, in the theory
${\cal L}={1\over2g_A}(\nabla\Phi)^2$. Notice that by tuning
$g={\ln 2\over 4\pi} g_A$,
we can match the coefficients in front of the logarithms.

We have shown that the RCS model yields a Gaussian
field theory, which shares (symbolically) the same correlation
functions as the ${\cal L}= {\ln2\over8\pi g}(\nabla\Phi)^2$ field theory.
The difference between the two theories is that the ultrametric
$d(x,y)$ is not the Euclidean metric $|x-y|$.
The scaling exponents for the two theories are nevertheless
the same, despite their different metrics.
We believe that the multifractal properties of the
critical wave function for Dirac fermions with random vector potential
can be obtained from the RCS realization. We give support to this
claim by comparing exact results for the RCS model with
numerical simulations for the ${\cal L}= {1\over2g^{\ }_A}(\nabla\Phi)^2$
theory.

We now turn to the calculation of $\tau(q)$ within the RCS model. It is
possible to define box probabilities on our random binary tree which
obey multifractal scaling of the form defined in Eq.
(\ref{exacttau(q)}). To this end, we define local random events at
level $n$ in the binary tree by
$
O^{\ }_n(x)=\exp\left[{2\Phi^{\ }_n(x)}\right]
$
[compare with Eq. (\ref{Diracprobdistr})].
{}From Eq. (\ref{Phin(x)}), we see that $O^{\ }_n(x)$ describes a
{\it random multiplicative process}. Be aware, however, that $O^{\ }_n(x)$
and $O^{\ }_n(y)$, for two points $x$ and $y$, are not independent
random variables.  From $O_n(x)$ we can construct another random
variable
\begin{equation}
Z^{\ }_n(q;\Phi^{\ }_n)=\sum_{j=1}^{2^n}O^q_n(x^{[j]}),
\label{Z_n(q)}
\end{equation}
where the summation extends over all microscopic boxes $\Omega^{\ }_j$
of volume $a^2=2^{-n}L^2$,
labelled by any $x^{[j]}\in \Omega^{\ }_j$.
In analogy to Eq. (\ref{exacttau(q)}), we then introduce
\begin{equation}
\tau^{\ }_n(q;\Phi^{\ }_n)\!=\!
-{2\over\ln2}
\left[
{\ln Z^{\ }_n(q;\Phi^{\ }_n)\over n}-q{\ln Z^{\ }_n(1;\Phi^{\ }_n)\over n}
\right],
\label{tau_n(q;O)}
\end{equation}
for any given random event $O^{\ }_n(x)=\exp[{2\Phi^{\ }_n}(x)]$.
Here, we had to divide
$O_n(x)$ by $Z^{\ }_n(1;\Phi^{\ }_n)$ to extract properly normalized
box probabilities: hence the second term in the bracket of Eq.
(\ref{tau_n(q;O)}).  Also, the prefactor 2 comes from space being two
dimensional, whereas the prefactor $\ln 2$ results from having chosen a
binary partitioning.
As a function of $q$ alone, $\tau^{\ }_n(q;\Phi^{\ }_n)$
shares by construction all the analytical properties of multifractal
scaling exponents for box probabilities \cite{Hentschel 1983}.  We are
going to show that the large $n$ limit of $\tau^{\ }_n(q;\Phi^{\ }_n)$
exists and, most importantly, is independent of the random event $O^{\
}_n$.  In other words, $\tau(q)=\lim_{n\rightarrow\infty}\tau^{\
}_n(q;\Phi^{\ }_n)$ is selfaveraging.

To this end, note that we can interpret $Z^{\ }_n(q;\Phi^{\ }_n)$ as
the (random) partition function of directed polymers on a binary tree.
The inverse temperature is $|q|$, there are $2^n$ possible distinct
configurations of directed polymers which are denoted by all paths
$(s^{\ }_1,\dots,s^{\ }_n)$ on the binary tree, and each configuration
is weighted by a {\it random} Boltzmann weight with random energy
$2\Phi^{\ }_n(x^{[j]})$.  It is clear from Eq.
(\ref{tau_n(q;O)}) that we are interested in the random variable
$ v^{\ }_n(q;\Phi^{\ }_n)=n^{-1}\ln Z^{\ }_n(q;\Phi^{\ }_n), $ which
is essentially the (random) free energy per unit length of the model
of directed polymers in a random medium. The asymptotic properties in
the limit $n\rightarrow\infty$ of the random variable $v^{\
}_n(q;\Phi^{\ }_n)$ have been investigated in \cite{Derrida
1988,Buffet 1993}.

It is shown in \cite{Buffet 1993} that {\it almost surely}
\begin{equation}
\lim_{n\rightarrow\infty}v^{\ }_n(q;\Phi^{\ }_n)=
\cases
{
\ln\left[\overline{Z^{\ }_1(q       ;\Phi^{\ }_1)}\right],
&$|q|\leq q^{\ }_c$,\cr
{|q|\over q^{\ }_c}
\ln\left[\overline{Z^{\ }_1(q^{\ }_c;\Phi^{\ }_1)}\right],
&$q^{\ }_c<|q|$.\cr
}
\label{asymptoticv_n}
\end{equation}
Averaging over the random variables $\phi^{\ }_{\pm}$
on the first generation of the tree is denoted by an overline.  Eq.
(\ref{asymptoticv_n}) holds for any disorder such that the {\it
annealed} average $\overline{Z^{\ }_1(q;\Phi^{\ }_1)}$ is a well
defined even function of $q$.  We have restricted ourselves to Gaussian
random variables.  The critical value $q^{\ }_c$ is defined by the
unique minimum of $q^{-1}\ln[\overline{Z^{\ }_1(q ;\Phi^{\ }_1)}]$,
$q\geq0$.  The existence of a critical ``temperature'' $q^{-1}_c$ was
anticipated in \cite{Derrida 1988}. Above the critical temperature, the
quenched and annealed averages over the free energy agree.  Below the
critical temperature, they do not. In Ref. \cite{Derrida 1988} the low
temperature phase is identified as a glassy phase.

The Gaussian average on the right hand side of Eq. (\ref{asymptoticv_n})
is easily performed. For example
\begin{equation}
q^{\ }_c= \sqrt{{\ln2\over2g}}.
\label{q^2_c}
\end{equation}
There are {\it two} distinct regimes depending on the strength $g$ of the
disorder.  In the weak disorder regime defined by $q^{\ }_c>1$, Eqs.
(\ref{q^2_c},\ref{asymptoticv_n},\ref{tau_n(q;O)}) yield almost surely
\begin{equation}
\tau(q)=
\cases
{
2(1-{{\rm sgn} q\over q^{\ }_c})^2q,&$q^{\ }_c<|q|   $,\cr
(2-{2\over q^2_c}q)(q-1), &$q^{\ }_c\geq|q|$,\cr
}
\label{weaktau(q)}
\end{equation}
in the limit $n\rightarrow\infty$. In the weak disorder regime, the
parabolic approximation $\tau^*(q)=D^*(q)(q-1)$, where
$D^*(q)=2-(2/q^2_c)q$, is exact for all moments $q$ satisfying
$|q|\leq q^{\ }_c$. Here, the parabolic approximation (PA) is obtained
from the {\it annealed} disorder average over Gaussian random variables $
\tau^*_n(q)\!=\!
$
$
-[{2/(2\ln2)}]
$
$
[
  \ln \overline{Z^{\ }_n(q;\Phi^{\ }_n)}
-q\ln \overline{Z^{\ }_n(1;\Phi^{\ }_n)}
]
$ instead of the quenched average implied by Eq. (\ref{tau_n(q;O)}) .
The PA breaks down for large moments in view of the inequality $
\lim_{n\rightarrow\infty}
n^{-1}\overline{\ln Z^{\ }_n(q;\Phi^{\ }_n)}
$
$
<\ln\overline{Z^{\ }_1(q;\Phi^{\ }_1)},
$
$
|q|>q^{\ }_c\geq1.
$
It should be noted that the field theory approach of
\cite{Mudry 1995} to Dirac fermions coupling
to the disorder of Eq. (\ref{Diracprobdistr})
relates primary fields with negative scaling dimensions to the multifractal
scaling exponents $\tau^*(q)$ (for integer valued $q$) with
$q^2_c=2\pi/g^{\ }_A$.  In the strong disorder regime defined by $q^{\
}_c\leq1$, the quenched and annealed averages of $Z^{\ }_n(q;\Phi^{\
}_n)$ are unequal for all integer moments. The PA completely breaks
down for integer moments $q$:
\begin{equation}
\tau(q)=
\cases
{
{4\over q^{\ }_c}(q-|q|),&$q^{\ }_c<|q|     $,\cr
-2(1-{q\over q^{\ }_c})^2, &$q^{\ }_c\geq|q|$.\cr
}
\label{strongtau(q)}
\end{equation}

The multifractal analysis of Eqs.
(\ref{weaktau(q)},\ref{strongtau(q)}) consists in performing a
Legendre transformation from $\tau(q)$ to $f(\alpha)$. The Legendre
transformation is well defined since $\tau(q)$ is concave in the weak
as well as strong disorder regimes.  The interpretation of $f(\alpha)$
is that it yields almost surely the Hausdorff dimensions of interwoven
Cantor sets characterizing a {\it typical} distribution
$\mu(x,\Phi^{\ }_{\rm typ})$.
Indeed, $f(\alpha)$ is strictly
positive on its domain of definition, in contrast to the Legendre
transform $f^*(\alpha)$ of the scaling exponents $\tau^*(q)$ (which we
recall are calculated from {\it annealed} disorder
averages as in \cite{f(alpha)negative}).
For weak disorder, $f(\alpha)$ is a parabola defined on $D^{\
}_+\leq\alpha \leq D^{\ }_-$, where $D^{\ }_{\pm}=2(1\mp1/q^{\
}_c)^2$.  For strong disorder, $f(\alpha)$ is a parabola defined on
$0\leq\alpha\leq 8/q^{\ }_c$. In both cases, the parabola takes the
maximum value 2, the dimensionality of space, and vanishes when
$\alpha=D^{\ }_{\pm}$ with {\it finite} slopes.  The finiteness of the
slope of $f(\alpha)$ at the end points of its domain of definition
comes about as a result of the transition to linear behavior of
$\tau(q)$ when $q\geq q^{\ }_c$. This transition is interpreted as a
phase transition to a glassy phase in the context of directed polymers
in random medium.  This property of $f(\alpha)$ in the RCS
construction should be contrasted to the deterministic two scale
Cantor set where the two slopes of $f(\alpha)$ are infinite at $D^{\
}_{\pm}$ \cite{Hentschel 1983,Halsey 1986}.

We have compared numerical calculations of the scaling exponents
$v(q,\Phi)$ obtained from the ${\cal L}= {\ln2\over8\pi g}(\nabla\Phi)^2$
field theory with the exact results derived from the RCS construction.
We generated four Monte Carlo realizations for the disorder
$\exp[-{\ln2\over8\pi g}\int d^2x(\nabla\Phi)^2(x)]$
on a square lattice made of $512\times512$ sites.
Within statistical fluctuations due to the finite size of the lattice,
the four Monte Carlo calculations $v^{\ }_{\rm MC}(q,\Phi)$
and $v(q)$ from the RCS model [Eq. (\ref{weaktau(q)})] agree.

Finally, we can obtain directly from
${\cal L}$$={1\over2g_A}$$(\nabla\Phi)^2$
the scaling behavior of the sequence of ratios
$R_m(q)$$=\overline{Z^m(q;\Phi)}$$/\overline{Z(q;\Phi)}^m$.
Here,
$Z(q;\Phi)$$=\int\! d^2x$$\exp[2q\Phi(x)]$, $m$ is an integer, and
averaging is defined by Eq. (\ref{Diracprobdistr}).
With the help of dimensional analysis, we find
that $R_m(q)$ contains a term scaling as $(a/L)^{\Delta_m(q)}$, with
$\Delta_m(q)=2(m-1)[1-{mg^{\ }_A\over2\pi}q^2]$.
Taking the thermodynamic limit yields a finite non-universal number
for $R_m(q)$ if $q<q_{cm}$, whereas $R_m(q)$ diverges if $q\geq
q_{cm}$.  Thus, there exists a {\it decreasing} sequence of critical
moments $
q^2_{cm}={2\pi\over mg^{\ }_A}.
$
It can be shown along the lines of \cite{Buffet 1993}
that by analytically continuing $q^{\ }_{cm}$ to the limit $m=1$, we find
the critical moment $q^{\rm Dirac}_c\equiv\sqrt{2\pi/g^{\ }_A}$,
which is the counterpart of Eq. (\ref{q^2_c}).
For $q<q^{\rm Dirac}_c$, we can commute disorder averaging with taking the
logarithm of $Z(q;\Phi)$ and thus rely on conventional
techniques of field theory to calculate $\tau^{\ }_{\rm Dirac}(q)$.
For $q\geq q^{\rm Dirac}_c$, disorder
averaging does not commute anymore with taking the logarithm,
and we have to use the RCS model to calculate $\tau^{\ }_{\rm Dirac}(q)$.
Notice that with
the identification $g^{\ }_A=[(4\pi)/\ln 2]g$,
the critical moments calculated in the two models
coincide.

All these arguments strongly suggest that Eq. (\ref{tau_n(q;O)})
describes the exact multifractal properties of the critical wave function
for Dirac fermions with random vector potential. Moreover,
our results support the conjecture on the universality of
$\tau(q)$, ({\it i.e.}, independence of impurity realizations) at
criticality in the problem of localization in two dimensions
\cite{Pook 1991}. Finally, the methods developed in this
letter may be just a first manifestation of more general principles.
We have found a faithful map between a particular RCS and a massless
free field theory. These two theories share symbolically the same
correlation functions, even though they are defined in different
metric spaces. Quantities such as scaling exponents should not depend
on the particular realization of the field theory, and thus one has
the choice to work with whichever is simpler.
Whether such maps between field theories in different metric
spaces can be generalized to non-trivial field theories is an
interesting open question.

We would like to thank M. Kardar and H. Orland for pointing out to us
Ref. \cite{Derrida 1988}.  This work was supported by NSF grants
DMR-9411574 (XGW) and DMR-9400334 (CCC).  CM acknowledges a fellowship
from the Swiss Nationalfonds and XGW acknowledges the support from
A.P. Sloan Foundation.

\vskip -0.5 true cm

\vfill

\end{multicols}
\end{document}